\documentclass{aa}  

\usepackage{graphicx}
\usepackage{multicol}

\usepackage{txfonts}
\usepackage[colorlinks=true,citecolor=blue,linkcolor=blue]{hyperref}
\usepackage{float}

\usepackage{natbib}
\bibpunct{(}{)}{;}{a}{}{,}
\defcitealias{andersson24starclusters}{A24}

\newcommand{\Msun}{{\rm M}_{\odot}}

\usepackage{etoolbox}
\makeatletter
\AtBeginDocument{
  \@ifpackageloaded{lineno}{\nolinenumbers\renewcommand{\linenumbers}{} }{}
}
\makeatother

\begin{document}

   \title{Stellar Feedback Effects on the Mass Distribution of Clouds and Cloud Complexes}

\titlerunning{Feedback Effects on Clouds}
\authorrunning{Quinalha et al.}
   \author{Luanna Veroneze Quinalha
          \inst{1}
          \and
          Eric P. Andersson \inst{2}
          \and 
          Mordecai-Mark Mac Low \inst{2}
          }

   \institute{Department of Physics and Astronomy, Barnard College, New York, NY, 10025, USA\\
              \email{lvq2101@barnard.edu}
         \and
             Department of Astrophysics, American Museum of Natural History, 200 Central Park West, New York, NY, 10024, USA
             }

   \date{Received ; accepted }
 
  \abstract  
   {Galaxy evolution is sensitive to the details of how stars inject feedback into their surroundings. In particular, the stellar feedback from star clusters strongly affects gas motions and consequently the baryonic cycle. More massive clusters have stronger effects. Our previous results show that the star cluster mass distribution in dwarf galaxies depends on feedback, as strong pre-supernova feedback, particularly ionizing radiation, results in fewer high-mass star clusters.}
   {We investigate the mass distribution of gas clouds in dwarf galaxies. Since star clusters form from the collapse of gas clouds, we expect a similar feedback dependence in both their mass distributions, so we hypothesize that pre-supernova feedback results in fewer high-mass gas clouds.}
   {To test our hypothesis, we use an isocontour analysis at three cutoff densities of $10, \ 10^{1.5} \mbox{, and } 10^{2} \mbox{ cm}^{-3}$ to identify gas clouds from 
    dwarf galaxy simulations performed with the \textsc{ramses} adaptive mesh refinement code.
    We calculate mass distributions for models that implement different combinations of the feedback modes: supernovae, stellar winds from massive stars, and ionizing radiation.}
   {We find that the mass distribution for clouds with $n > 100 \mbox{ cm}^{-3}$ is independent of feedback, but the mass distribution for cloud complexes with $n > 10\mbox{ cm}^{-3}$ is more top-heavy in the presence of radiation. Winds do not affect the mass distribution at any scale studied.}
   {This contradicts our hypothesis that the mass distribution of gas clouds would show similar feedback-dependence to the mass distribution of star clusters. Instead, our results show no feedback dependence in the mass function of dense clouds with $n > 100 \mbox{ cm}^{-3}$, suggesting their mass distribution is predominantly set by gravity. We conclude that the shape of the star cluster mass function must be determined by a combination of intra-cloud feedback regulation of star formation (i.e., regulation of star formation within a cloud due to feedback of stars formed from that cloud itself) and, in the case of radiation, effects on the temperature of the parent gas clouds.}


   \maketitle

\section{Introduction}\label{intro}

A long-standing question about the star formation process is the connection between star clusters and the clouds from which they emerge \citep{Lada&Lada2003, Dobbs+2014, Krumholz+2019, Chevance2023, Knutas+2025, Pedrini+2025}. Because of the timescales involved, this question is currently impossible to address directly through observations of a single cloud evolving into a star cluster---statistical studies of populations are instead required. Star formation is often investigated through theoretical models constrained by accessible observables. For example, the depletion time, the rate at which gas is consumed by star formation, is accessible empirically by comparing the available molecular gas mass and star formation rate. While this provides insight into the star formation process, large scatter is observed in the depletion time, particularly at small scales \citep{Onodera+2010, Schruba+2010}. Another property guiding our understanding of these processes is the star formation efficiency per free-fall time, typically observed to be small \citep[only a few per cent;][]{Krumholz+2019}. However, while convenient from a modeling perspective, this property relies on the assumption that there exists a scale where interstellar gas is converted into stars through the collapse of a single structure in free-fall. In reality, the interstellar medium is filamentary and supported by turbulence \citep{MacLow&Klessen2004, McKee&Ostriker2007, Burkhart2018}, has dynamics affected by magnetic fields \citep{Crutcher+2012}, and is affected by large-scale shear \citep{Dobbs+2014, Jeffreson+2020}. While it is possible to attribute the low efficiency of star formation to these factors, it might also be the case that stars originate from gas that is approximately in this state, but it happens rarely but rapidly. Therefore, when viewed as an average over a larger scale, star formation appears inefficient \citep{Semenov+2017}.

Rather than looking at individual objects, populations of clusters and clouds can be analyzed, allowing inference of the evolution from the distribution of objects at different stages \citep{Knutas+2025}. Particular attention has been drawn to the similar shape of the mass distribution of these clusters and clouds \citep[see, e.g.,][]{Guszejnov+2018}. For star clusters, the mass distribution typically shows a power-law function with a steep ($c\simeq -2$) negative slope \citep{PortegiesZwart+2010, Krumholz+2019, Adamo+2020, Pedrini+2025}. Interestingly, the mass distribution for gas structures shows a similar profile \citep{JCMT}, although there is some dependence on environment \citep{Rosolowsky2005,Colombo_2014,rosolowsky2021}.  

A major challenge to addressing this question is the fact that clouds are part of a continuum of structures in the interstellar medium, so individual cloud properties such as mass are hard to define \citep[as highlighted in the review by][]{Chevance2023}. Furthermore, these properties are often derived from structures determined by different observables that can show different morphologies. For example, molecular clouds observed through CO line emission often appear as structures with well-defined boundaries, while coincident structures in dust extinction maps show connected filaments as part of larger complexes \citep[see][for detailed discussion and examples]{Chevance2023}.

\begin{figure}
    \centering
    \includegraphics[width=\linewidth]{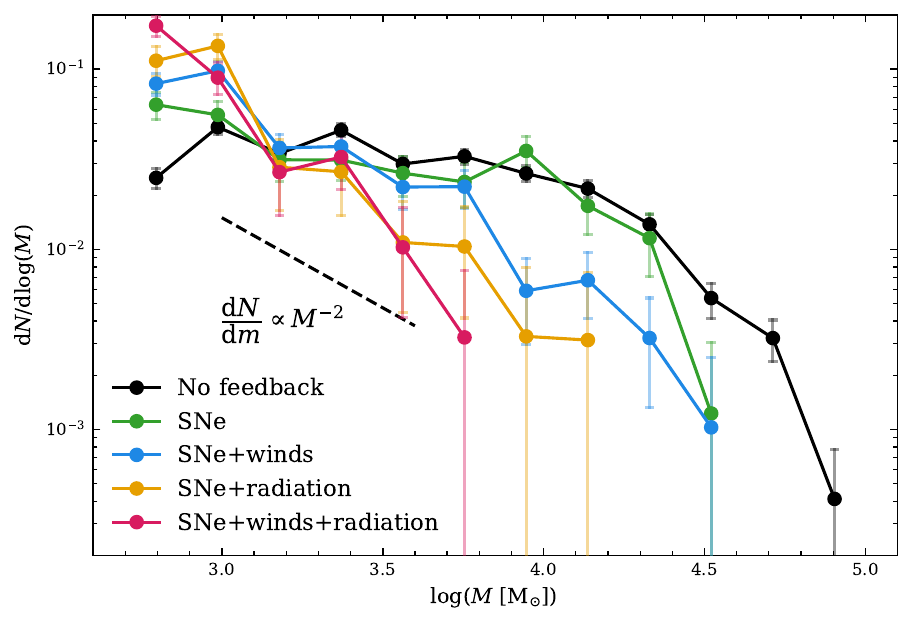}
    \caption{Cluster initial mass function, normalized to the total number of clusters (defined as stellar groups with ages $<25$~Myr) found in each simulation after 300 Myr. The error bars indicate the 16th and 84th percentiles computed using bootstrapping. The dashed line shows the slope of scale-free formation models, often associated with the initial cluster mass function (from \citetalias{andersson24starclusters}, where more details are provided).}
    \label{fig:cluster}
\end{figure}

The goal of this work is to determine how the shape of the cloud mass distribution varies as a function of feedback in hydrodynamical simulations of galaxies. We compare this to the mass distribution of star clusters derived from the same simulations \citepalias[presented in][]{andersson24starclusters} and replicated in Figure \ref{fig:cluster}. Of particular interest is that different sources of feedback were included in simulations of the same galaxy, while the remaining model parameters were kept the same. In this way, we address: 1) how different feedback mechanisms determine the cloud mass distribution; 2) how sensitive the star cluster mass distribution is to the cloud distribution from which it originates.

For our simulations, \citet[][hereafter \citetalias{andersson24starclusters}]{andersson24starclusters} showed that the star cluster mass distribution depends on feedback, transitioning from a wide function approaching a log-normal shape in the absence of feedback, to a steep power-law function when sources of feedback prior to supernovae (SNe), particularly ionizing radiation, are active. Furthermore, this effect was shown to be tightly linked to the timescale of the active feedback, including the details that affect this timescale, e.g., the main-sequence lifetime for the most massive SN progenitor assumed. We test whether this effect comes from similar behavior in the cloud mass distribution, or whether it is related to the formation of individual star clusters inside the clouds themselves.

We start by detailing our simulations and cloud selection algorithm in Section \ref{methods}, followed by a presentation of our main results in Section \ref{results}. In Section \ref{discussion}, we interpret the effects of feedback on cloud and cloud complex mass distributions, and briefly discuss other computational and observational data. Section \ref{conclusion} summarizes our work.

\section{Method}\label{methods}

\subsection{Simulations}
The galaxy simulations used in our analysis are described in \citetalias{andersson24starclusters}. Briefly summarized, all simulations are executed with \textsc{ramses-rt} \citep{2013MNRAS.436.2188R, 2015MNRAS.449.4380R}, a radiative transfer extension of the adaptive mesh refinement and $N$-body code \textsc{ramses} \citep{2002A&A...385..337T}. Heating and cooling include the equilibrium cooling from heavy elements implemented in \textsc{ramses}, as well as non-equilibrium ionization of H and He coupled to the radiation solver following the method described in \citet{2013MNRAS.436.2188R, 2015MNRAS.449.4380R}, which relies on an M1 closure for the Eddington tensor. In addition, we track the evolution of H$_2$ following the method of \citet{Nickerson+2018}. The refinement criterion employs a quasi-Lagrangian technique aiming for approximately eight particles in each cell and cell refinement at a mass threshold of $800\,{\rm M_{\odot}}$. The cell size is 3.6 pc at the maximum refinement level. The simulations ran for 1000 Myr with an output cadence of 25 Myr (see \citetalias{andersson24starclusters} for visualizations of, e.g., star formation history).

The initial conditions mimic an isolated dwarf galaxy using the tool \textsc{MakeDiskGalaxy} \citep{2005MNRAS.364.1105S}. Embedded in a dark matter halo following a \citet{navarro1997} profile is a disc with gas mass $M_{\rm g,disc}=7\times 10^{7}\,{\rm M_{\odot}}$ and stellar mass $M_{\rm s,disc}=10^{7}\,{\rm M}_{\odot}$ and an exponential radial density profile having scale length $1.1\,{\rm kpc}$. The gas has an initial temperature of $10^{4}$ K. The initial metallicity of the gas is set to 0.1 $\rm Z_{\odot}$.

Stars and dark matter are tracked by particles of masses $100\,{\rm M}_{\odot}$ and $10^{4}\,{\rm M}_{\odot}$, respectively. Star formation proceeds by forming star particles in cells with gas density $n>10^{3}\,{\rm cm}^{-3}$ and temperature $T<10^{4}$~K at each time step. In cells exceeding these thresholds, a discrete number of single stellar population particles, each with mass $100\,{\rm M}_{\odot}$, can be spawned. The number of star particles is drawn from a Poisson function with the mean number of stars given by the cell gas mass divided by a star formation rate \begin{equation}
    \dot{\rho}_{\star}=\epsilon_{\rm ff}\frac{\rho_{\rm g}}{t_{\rm ff}},
\end{equation}
multiplied by the timestep. We determine the star formation rate by assuming an efficiency $\epsilon_{\rm ff} = 0.1$ per local gas free-fall time $t_{\rm ff}=[3\pi/(32{\rm G}\rho_{\rm g})]^{1/2}$. When regulated by feedback, this star formation algorithm was shown to reproduce the star formation relation \citep{Agertz+2016}, observed star formation efficiencies \citep{Grisdale+2019}, as well as scaling relations for clouds \citep[e.g.,][]{Larson1981} in the Milky Way \citep{Grisdale+2018}.

The stellar feedback model includes stellar winds and SNe following the method of \citet{Agertz+2013}, while radiation is treated through the radiative transfer method \citep{2013MNRAS.436.2188R}. All feedback is calibrated assuming the initial mass distribution from \cite{2001MNRAS.322..231K}. Fast stellar winds injected by massive ($M_*>8\,{\rm M}_{\odot}$) stars are implemented as a source of momentum with a wind velocity of $v_w = 1000\mbox{ km s}^{-1}$ calibrated using \textsc{starburst99} \citep{Leitherer+1992, Leitherer+1999}. Core-collapse SNe are stochastically sampled as discrete events using the main sequence lifetimes from \cite{1996A&A...315..105R}, with progenitors in the mass range $M_*=8-30\,{\rm M}_{\odot}$, each injecting $E_{\rm SN} = 10^{51}$~erg of energy and momentum equivalent to 12~M$_{\odot}$ at a velocity of $3000\mbox{ km s}^{-1}$. Type Ia SNe follow the time-delay distribution of \citet{Maoz&Graur2017} assuming a time delay of 38 Myr and rate of $2.6\times10^{-13}\mbox{ yr}^{-1}M_{\odot}^{-1}$. We assume a mass loss of $1.4\,{\rm M}_{\odot}$ and a momentum release equivalent to $E_{\rm SN}$. Finally, radiation feedback includes scattered infrared photons (non-thermal), direct radiation pressure from optical and far-ultraviolet, Lyman-Werner radiation for H$_2$ dissociation, and photoionizing radiation around the ionizing energies of \ion{H}{ii}, \ion{He}{i}, and \ion{He}{ii} (bands defined as in \citealt{Agertz+2020}).

\subsection{Cloud selection}\label{sec:cloud_selection}
\begin{figure*}
    \centering
    \includegraphics[width=\linewidth]{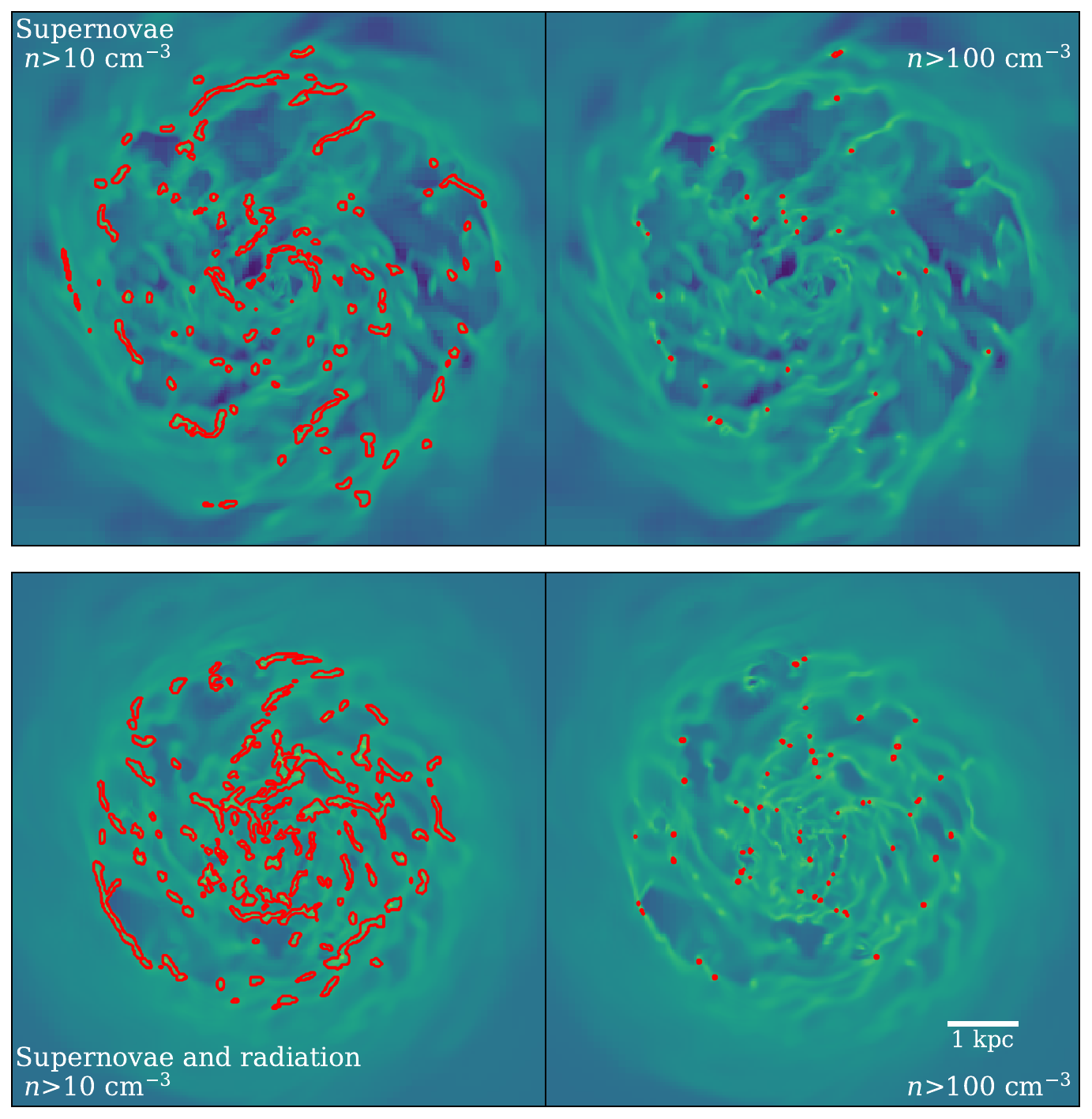}
    \caption{Root node contours in projected gas density plots for different simulations (SNe in the top row, SNe+radiation in the bottom row) and density thresholds ($n >$10 cm$^{-3}$ in the left column, and $n>10^2$~cm$^{-3}$ in the right column). As expected, astrodendro selects fewer and smaller structures at a higher density cutoff.} 
    \label{fig:contours_panel}
\end{figure*}

To find clouds, we first transfer the \textsc{ramses} data to a uniform grid at the maximum resolution level. We omit data from simulation outputs in the first 300 Myr to avoid spurious effects from the initial conditions. We then employ an isocontour analysis on the density data of each simulation output using the astrodendro library \citep{2019ascl.soft07016R} with three different density cutoffs: $n > 10 \mbox{ cm}^{-3}$, $10^{1.5}\mbox{ cm}^{-3}$ and $10^2\mbox{ cm}^{-3}$. We study the properties of the largest-scale root nodes in the dendrogram for each of the three cutoffs, which we call cloud complexes, big clouds, and clouds.  These choices will be further discussed in the next section, but for the moment, contrasting the right and left sides of Figure \ref{fig:contours_panel} provides some visual intuition for our nomenclature. By using astrodendro for our isodensity analysis, we obtain information about cloud and cloud complex substructure that is being used to understand simulations  \citep{Torch1, Torch2} of star cluster formation from the clouds \citetext{Jiang et al., in prep}.

\section{Results}\label{results}

\begin{figure}
    \centering
    \includegraphics[width=\linewidth]{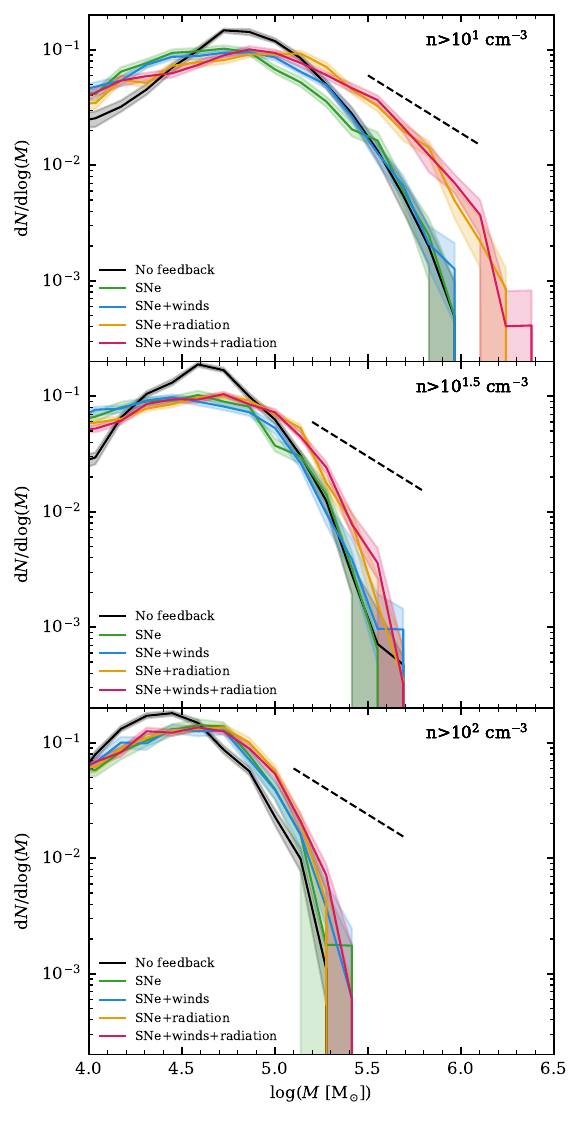}
    \caption{ 
       Mass distributions for cloud complexes with $n>10$~cm$^{-3}$ (top), big clouds with $n>10^{1.5}$~cm$^{-3}$ (middle), and clouds with $n>10^2$~cm$^{-3}$ (bottom). The lines show the median values, while the shaded regions show the 16th--84th percentile range. All distributions shown contain structures from multiple snapshots, where consecutive snapshots are 25 Myr apart.
    Dashed line shows ${\rm d}N/{\rm d}\log(M) \propto M^{-1}$, i.e., ${\rm d}N/{\rm d}M \propto M^{-2}$.} 
    \label{fig:massfunction}
\end{figure}

We focus on gas structures selected in three-dimensional space using density contours, as described in Section~\ref{sec:cloud_selection}. Figure \ref{fig:massfunction} shows the mass distribution obtained for different gas structures, plotting the medians and 16th--84th percentile ranges. \footnote{We calculate statistical properties using the bootstrapping technique, drawing 100 samples of 80\% of the total data in each case.} We note that gas in the simulation without feedback shows significantly different behavior compared with the other models, so we do not focus on it in Figure~\ref{fig:massfunction}, but rather on comparing the other simulations, keeping the simulation without feedback as a reference. Our mass resolution on the refined grid is under $800 \rm \, \Msun$, while the peaks of our mass distributions are over an order of magnitude higher, and thus probably resolved numerically. 

For cloud complexes (top panel in Figure~\ref{fig:massfunction}), all simulations show similar fractions of low mass ($M\lessapprox 10^{5} \,\Msun$) objects. This changes at larger masses, where the mass distributions for simulations including radiation peak at a larger mass and have a shallower slope at high masses, producing a larger number of massive clouds, compared to the mass distributions for simulations without radiation.  The distribution of cloud volumes behaves similarly, while the distribution of average cloud densities is mostly feedback-independent (see the Appendix~\ref{appendix}). This suggests that the larger number of more massive cloud complexes formed in the presence of radiation corresponds to a larger number of larger cloud complexes. Our results thus suggest that radiation causes larger, more massive cloud complexes to form.  Figure~\ref{fig:contours_panel} illustrates this behavior---the low-density contours delineate larger, less fragmented structures when radiation is included in the simulation, a structure that remains consistent throughout the simulation. 

For clouds (bottom panel in Figure~\ref{fig:massfunction}), the low mass end ($M<10^{4.5}\,{\rm M}_{\odot}$) is similar to that for the cloud complexes. However, at higher masses, the distribution function is significantly steeper. Furthermore, the cloud mass distributions in the different simulations overlap within the shown percentiles, indicating that there is insignificant feedback dependence. 

For big clouds, with masses and bounding densities between cloud complexes and clouds, we see a transition where radiation plays a small role at the massive end and no role at the low mass end. This suggests an inverse proportionality between density cutoff and radiation dependence of the mass distribution, at least within the 10--100~cm$^{-3}$ range. 

At the massive end of the mass distribution, its slope is very steep. We include the ${\rm d}N/{\rm d}M\propto M^{-c}$ with $c=2$ derived from empirical estimates \citep{Chevance2023} as a dashed line to guide the eye. However, keep in mind that our mass distributions do not display a log-normal shape overall, consistent with other simulation work, which tends to fit a power-law only on the high-end of the spectrum \citep[see, e.g.,][]{Renaud2024, HOPexample, FOFexample}. More generally, the lower-mass cloud distribution with $M\lesssim10^{4.5}\,{\rm M}_{\odot}$ remains similar regardless of density selection in all simulations, while the distribution of more massive structures more rapidly steepens when considering higher density gas.

\begin{figure*}
    \centering
    \includegraphics[width=\linewidth]{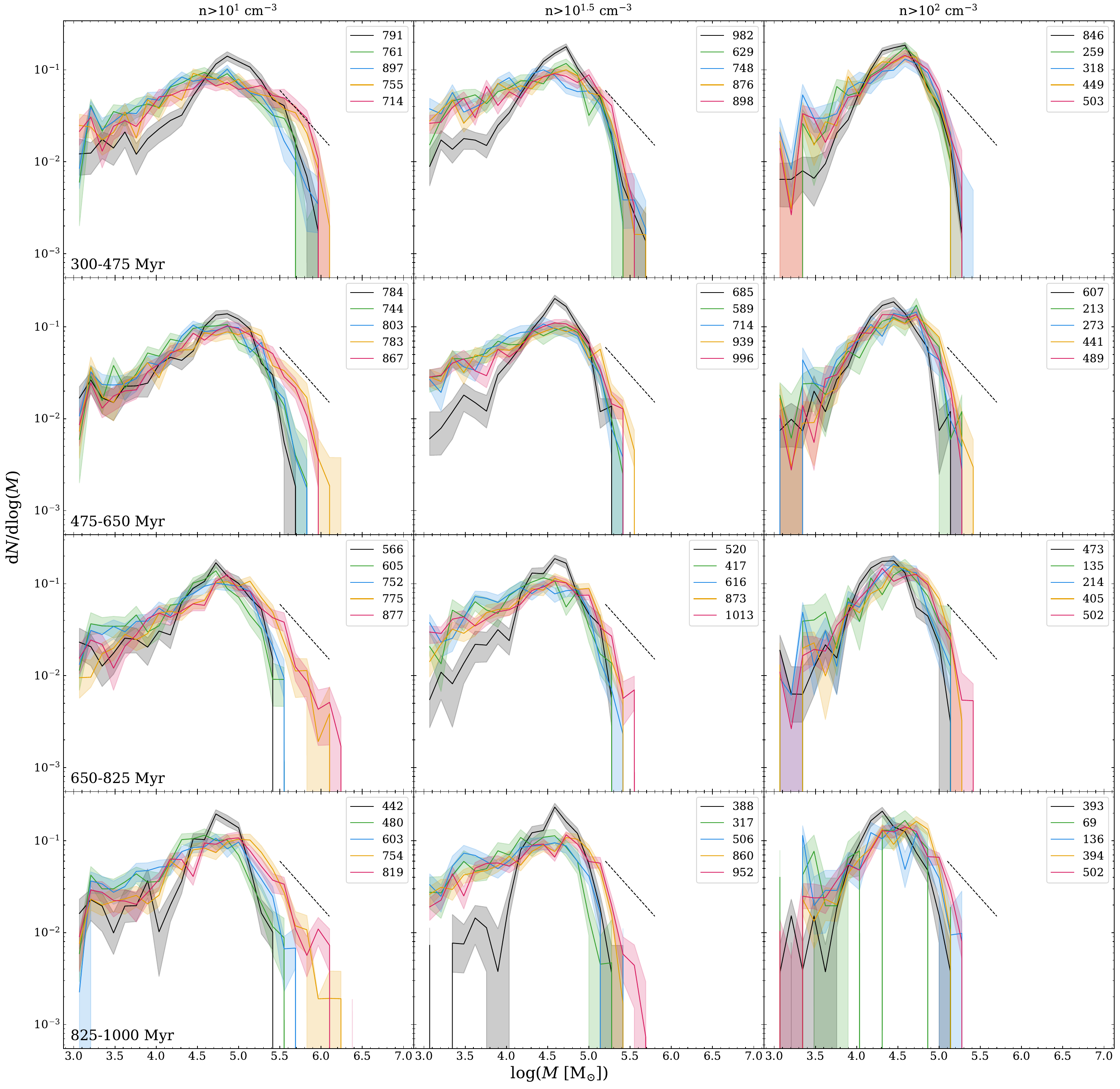}
    \caption{Cloud mass distributions in 175 Myr time bins. Feedback is encoded through color as in Figure \ref{fig:massfunction}, while the number of clouds in each distribution is given in the legend. The lines show the median values, while the shaded regions show the 16th–84th percentile range.} 
    \label{fig:mass-func-time-dependence}
\end{figure*}

Different simulations had different numbers of clouds, as shown in Figure~\ref{fig:mass-func-time-dependence}. In particular, adding winds and adding radiation both lead to an increase in the number of structures (clouds, big clouds, and cloud complexes) at all three density cutoffs. However, there does not seem to be a differential effect for clouds ($n>10^2$~cm$^{-3}$): the number of clouds is changed at all masses, leaving the cloud mass spectrum unchanged. 

The mass distributions of clouds selected with $n>10^{1.5} \, \rm cm^{-3}$ and $n>10^{2} \, \rm cm^{-3}$ are mostly time independent (Figure~\ref{fig:mass-func-time-dependence}), but for cloud complexes ($n>10 \mbox{ cm}^{-3}$), the mass distribution changes with time. In particular, the difference between simulations with and without radiation at the high end of the mass distribution is more dramatic later in the simulations (see Figure~\ref{fig:mass-func-time-dependence}).

\begin{figure}
    \centering
    \includegraphics[width=\linewidth]{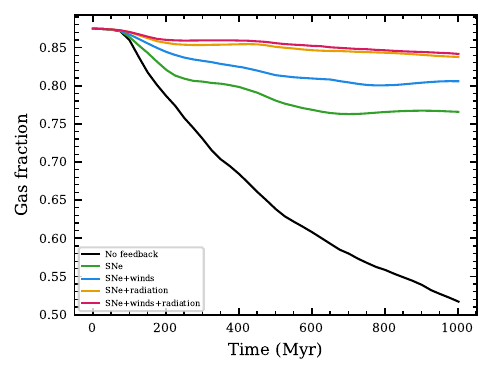}
    \caption{Gas fraction $f_{\rm gas}$ (Eq.~\ref{eq:gas-frac}) throughout time for all simulations as shown in the legend.}
    \label{fig:gasfractions}
\end{figure}

\begin{figure*}
    \centering
    \includegraphics[width=\linewidth]{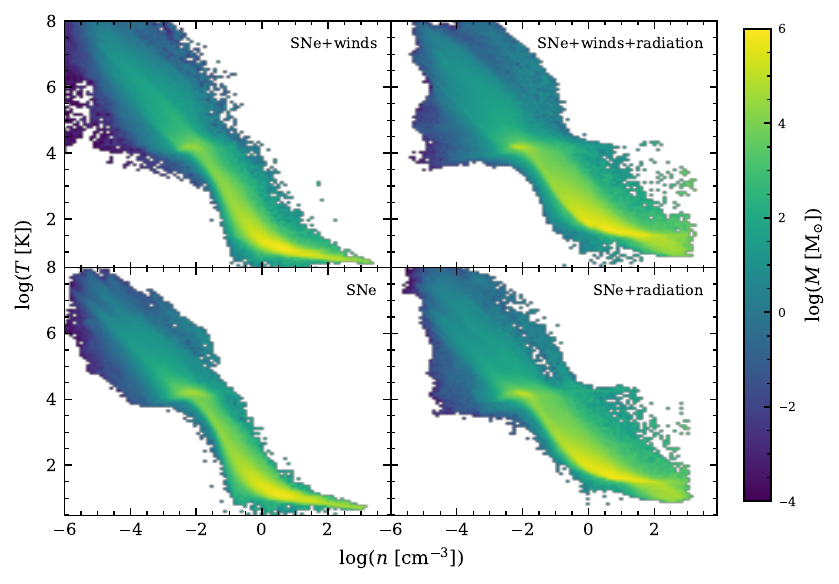}
    \caption{Temperature as a function of density, color coded by mass, for all gas in each simulation at time 750 Myr.} 
    \label{fig:phasediagrams}
\end{figure*}

\section{Discussion}\label{discussion}

\subsection{Why radiation impacts cloud complex mass function}\label{subsec1}

Radiation impacts a galaxy's gas fraction 
\begin{equation}
f_{\rm gas} = M_{\rm gas} / (M_{\rm gas} + M_{\rm stars}). \label{eq:gas-frac}
 \end{equation}   
Figure \ref{fig:gasfractions} shows the time evolution of gas fraction in the five simulations. Winds and radiation both lead to a diminished decrease in the gas fraction, although when radiation is present, the impact of adding winds is much less significant. Removal of gas through star formation alone cannot explain the evolution in gas fraction for the simulations with feedback, since star formation is higher before $600\,{\rm Myr}$ in simulations without radiation and lower after \citepalias[see Figure~2 in][]{andersson24starclusters}. Therefore, gas flows in and out of the disk must play a role. Radiation has been shown to decrease gas inflows and outflows \citep[e.g.,][]{Smith+2021}, which appears to explain the discrepancy. A higher gas fraction leads to more gas available to form cloud complexes, which is consistent with the larger number of structures found in simulations with winds, radiation, or both (Figure~\ref{fig:mass-func-time-dependence}). However, while introducing winds simply leads to more cloud complexes, introducing radiation also changes their mass distribution, making it more top-heavy. While \cite{Renaud2024} did find that increasing gas fraction from $f_{\rm gas} = 0.1$ to $0.4$ altered the shape of the clump mass distribution, they tested significantly higher changes in gas fraction without altering feedback modes, and using a different cloud extraction method (see Section~\ref{subsec4} for details).

The thermal state of the gas plays a key role in determining the density structure. Because introducing radiation changes the balance between cooling and heating, we expect it to be responsible for the effect we find in the distribution of cloud complexes. Indeed, density-temperature phase diagrams (Figure~\ref{fig:phasediagrams}) show that introducing radiation causes the gas equilibrium curves to shift up by about 0.5 dex in temperature. The Jeans mass is proportional to temperature, so higher temperatures allow larger masses to collapse without fragmentation \citep{Jeans1902}. This is consistent with our finding that radiation leads to higher mass cloud complexes (Figure~\ref{fig:massfunction}). Furthermore, radiation leaves more mass in warm ($T =10^2$--$10^4\,{\rm K}$), dense ($n >1\,{\rm cm}^{-3}$), high-pressure gas (Figure~\ref{fig:phasediagrams}).
\citetalias{andersson24starclusters} found that radiation leads to fewer and smaller star clusters, which could also explain why cloud complexes can grow larger before being disrupted by cluster feedback. The left-hand side of Figure \ref{fig:contours_panel} illustrates how radiation can reduce fragmentation in a galactic disk, resulting in more elongated cloud complexes (this is shown quantitatively by the cloud complex volume distribution in Figure~\ref{fig:volfunction}). Investigating cloud lifetimes is beyond the scope of this manuscript, but should be done in future work. It is worth noting that adding radiation changes the cloud complex mass distribution significantly, even compared to removing feedback entirely. While this may seem to imply that SNe and winds do not impact the cloud complexes formed, a closer look at their volume (Figure~\ref{fig:volfunction}) and density (Figure~\ref{fig:rhofunction}) distributions reveals that, compared to a galaxy with no feedback, one with SNe alone or SNe and winds together forms larger but less dense cloud complexes. The impact of SNe and winds thus seems to conspire in such a way as to leave the cloud complex mass function unaffected.

\subsection{Mass-velocity dispersion relation}\label{subsec2}

Figure \ref{fig:vdisp} shows the distribution of velocity dispersion against mass for cloud complexes, big clouds, and clouds. The leftmost panel shows that galaxies with radiation have cloud complexes reaching higher masses and velocity dispersions than simulations with only SNe, winds, or both. This is consistent with the radiation dependence of the cloud complex mass function (see Section~\ref{results}). The middle panel shows the same trend for big clouds, except that the difference is much smaller, again reflecting their mass distribution.

The rightmost panel of Figure~\ref{fig:vdisp} shows that the mass-velocity dispersion relation of high mass structures is independent of the type of feedback, but introducing a form of feedback affects this relation (c.f.\ the simulation without any feedback). Low mass objects have slightly suppressed velocity dispersion when exposed to radiation feedback. If one interprets the velocity dispersion as a proxy for energy, it may seem surprising that feedback has so little effect. However, what we change when we add a mode of feedback is the amount of energy injected per star throughout the simulation, not necessarily the total energy budget. Our results are consistent with a self-regulating system: as the energy per star increases, the number of massive clusters decreases \citepalias{andersson24starclusters}, likely maintaining a similar total feedback budget.

Finally, the no feedback case shows structures with higher velocity dispersion than their counterparts of the same mass in other simulations. This effect is exacerbated at higher density cutoffs, being most prominent for clouds. The excess velocity dispersion is likely due to gravity, now unresisted by feedback \citep{ibanez-mejia2016}, and more frequent collisions and mergers. This is consistent with the volume and density distributions in Figures \ref{fig:volfunction} and \ref{fig:rhofunction}, which show that structures identified in the galaxy with no feedback are smaller and denser than structures in other simulations.

\subsection{Feedback independence of cloud mass function and implications}\label{subsec3}

The gas cloud ($n>100$~cm$^{-3}$) mass distribution is feedback-independent (Sect.~\ref{results}). Feedback does affect the thermodynamics of the galaxy, as seen in Figure \ref{fig:phasediagrams}. It also changes the number of clouds in the galaxy, but without changing the shape of the distribution, as shown in Figure \ref{fig:mass-func-time-dependence}. This suggests the distribution of cloud masses is primarily dominated by gravity rather than feedback.
\citetalias{andersson24starclusters} found that galaxies with more feedback modes had fewer massive star clusters (see Fig.~\ref{fig:cluster}). They showed that the formation of star clusters is sensitive to the timing of the onset of stellar feedback. As such, they concluded that the immediate regulation of local star formation provided by radiation and stellar winds prevents the most massive clusters from forming. For brevity, we call this intra-cloud feedback regulation.

The simplest competing explanation would be that the feedback dependence of the star cluster mass function is a reflection of feedback effects on the clouds from which they originate. Our results show no feedback dependence in the mass distribution of clouds, contradicting this alternative.
\begin{figure*}
    \centering
    \includegraphics[width=\linewidth]{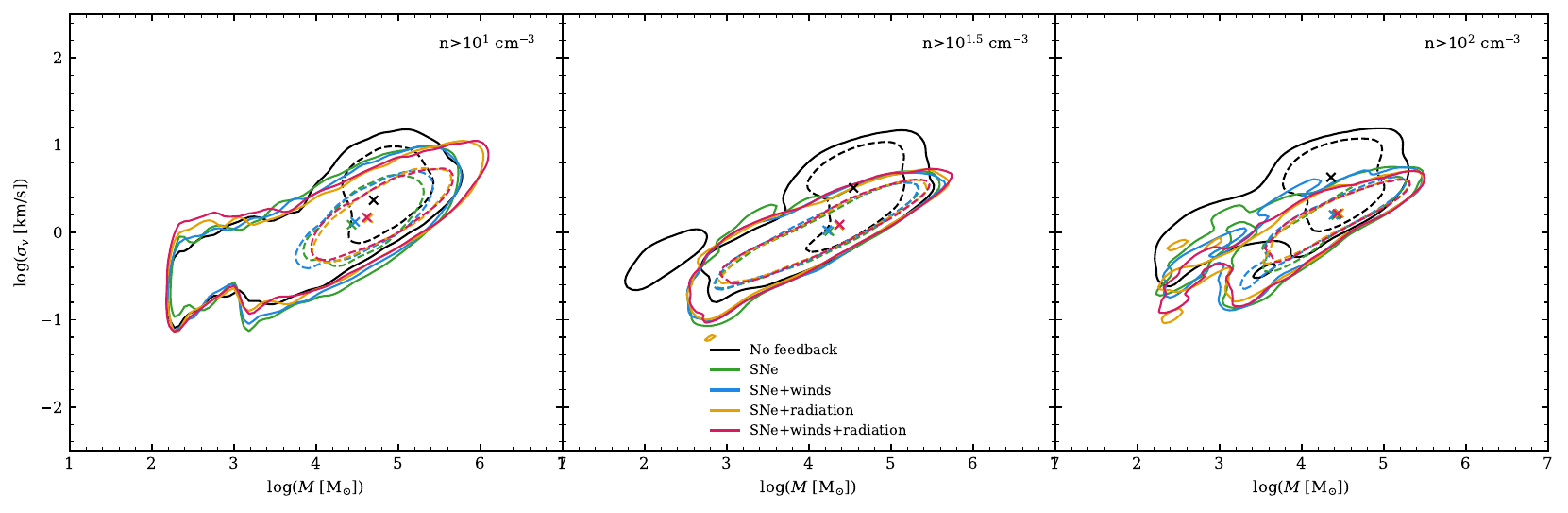}
    \caption{Distribution of velocity dispersion against mass for cloud complexes (left), big clouds (center), and clouds (right). The median of the distribution is marked with a cross, while the 68th and 95th percentiles are shown as dashed and solid lines, respectively. Different models are color-coded as indicated in the legend.}
\label{fig:vdisp}
\end{figure*}
A subtler alternative hypothesis is that feedback affects the thermodynamics or turbulence of the ISM, or both, in such a way as to prevent massive cluster formation. Figure \ref{fig:vdisp} shows the distribution of velocity dispersion against mass for clouds, big clouds, and cloud complexes. For clouds, although there is a dependence on the presence or absence of any feedback, all galaxies with feedback have similar distributions, suggesting that the impact of feedback on turbulence is also not responsible for the observed star cluster mass distributions.

Finally, the density-temperature phase diagrams in Figure~\ref{fig:phasediagrams} show that introducing radiation causes the gas equilibrium curves to shift up by about 0.5 dex in temperature at the cold end. Radiation leads to warmer clouds with higher thermal pressure supporting against collapse, likely explaining why \citetalias{andersson24starclusters} found that including radiation resulted in fewer massive star clusters. However, while winds were found to have the same effect on the star cluster mass function \citepalias{andersson24starclusters}, they do not significantly alter the temperature of gas clouds (top left panel of Figure~\ref{fig:phasediagrams}).

Thus, through a process of elimination, we agree with the argument of \citetalias{andersson24starclusters} that intra-cloud feedback regulation of star formation leads to fewer massive clusters in galaxies with winds and radiation, but we specify that the impact of radiation on gas cloud temperature appears to be the most important effect that we have studied.

\subsection{Comparison to other work}\label{subsec4}

In Section~\ref{subsec1} we mentioned the gas fraction results of \citet{Renaud2024}, but noted their different cloud extraction method. While we use isocontours in three dimensions (see Section\ \ref{methods}), they define circular shells around gas overdensities in two-dimensional projected simulation data. Additionally, their assumed initial gas fractions are significantly lower than the initial gas fraction in our galaxy. Finally, we simulate a galaxy with significantly lower mass compared to \citet{Renaud2024}. However, it may be that the effects of feedback found in our simulations would be different in a galaxy with a different initial gas fraction. Controlling for both gas fraction and modes of feedback could shed new light on how similarly or differently the effects of feedback on cloud dynamics manifest in different galactic environments.

\citet{Renaud2024} is only one example of how other work using simulations defines clouds differently than we do. Other cloud extraction algorithms include friends-of-friends, which dictates that two particles or grid cells with a density above the required threshold belong to the same group if they are within a specified distance from each other \citep[see, e.g.][]{FOFexample}; and HOP, which identifies clumps by linking every cell to its closest density peak and applying a number of regrouping criteria \citep[see, e.g.][]{HOPexample}. Not only are there many cloud extraction algorithms, but each method typically requires multiple parameters, which implies an even larger number of possible cloud definitions.

Using HOP with a density threshold of $30\,{\rm cm}^{-3}$ and a minimum peak density of $60\,{\rm cm}^{-3}$, \citet{HOPexample} found mass distributions with a power law slope close to $-1$ for six different simulations, while our data show a slope steeper than $-2$ at all density cutoffs explored (see Section~\ref{results}). The large discrepancy in our results may occur for a variety of reasons. Their analysis was made on simulations of stratified ISM boxes and galactic disks with significantly higher total mass and lower gas fractions than those of our dwarf galaxy. As such, our results may point to a true physical difference in cloud properties between these environments. However, there exists evidence of differences in clouds extracted with different methods  \citep[see, for example,][]{methodcomp1,methodcomp2,methodcomp3}. In fact, \citet{HOPexample} acknowledge that their mass distribution slope differs from other numerical and observational works and show that the discrepancy can originate from the extraction method used. Note, however, that we agree with their conclusion that the properties of clouds are not very sensitive to changes in physics (e.g., different thermodynamics in different codes) or to thermal pressure. We find that changing feedback can alter cloud temperatures but does not change the cloud mass distribution, so while we do not test different codes, we agree that the cloud mass distribution is not sensitive to changes in physics, including thermodynamics. Re-analysis of our data with different cloud extraction algorithms would both test the robustness of our results and facilitate comparison to previous work, but is outside the scope of this paper.

In addition to the challenge of how different works define different clouds, theoretical models for cloud and star formation also rely on different numerical models. For example, a limitation of our model is that the mass of star particles is so low at $100\,{\rm M}_{\odot}$ that they do not fully sample the initial mass function with each particle \citep[see, e.g.,][]{Krumholz+2015, Smith2021, Chevance+2022}. For SNe, this is accounted for by stochastically sampling discrete events; however, it remains a limitation for radiation and stellar winds that are treated as averages of a stellar population with a fully sampled initial mass function. This implies that for low-mass clusters with only a few particles, stellar feedback is likely overestimated. To assess the impact of this limitation, our results must ultimately be tested with a star-by-star approach \citep[see, e.g.,][]{Emerick+2018,Steinwandel+2022,Andersson+2023,Lahen+2023,Andersson+2025,Brauer+2025}. While star-by-star models treat the scatter introduced when sampling only a few massive stars per particle, these models typically assume a universal initial mass function. On the contrary, low mass star clusters likely have a systematically lower upper limit on the initial mass function compared to more massive clusters \citep{Weidner+2010, Grudic+2023}, which severely impacts star formation \citep{Lewis+2023}.

Finally, the main drawback to our three-dimensional analysis is that it is not well-suited for comparison with observed clouds. In Figures~\ref{fig:massfunction} and~\ref{fig:mass-func-time-dependence}, the dashed lines show the empirically derived \citep[e.g.,][]{Chevance2023} value of $c\approx-2$ for the power-law exponent in the mass distribution. There are few observational results specific to clouds in dwarf galaxies, but we note that \cite{ALMA_TDG} found a shallower slope of $-1.76\pm0.13$ for tidal dwarf galaxies. Most observational studies that have contributed to establishing the expected value of $c=-2$ used CO emission lines to delineate clouds and estimate their masses. Mass estimation relies on assuming a $^{12}$CO(1-0)-to-H$_{2}$ conversion factor, which depends on environmental conditions \citep[e.g.][]{glover2011} but is often assumed to be constant \citep[see, e.g.,][]{JCMT,ALMA_TDG}. Since our objective is to investigate gas cloud properties in dwarf galaxies as a complement to our previous studies about star clusters, we leave synthetic observations and comparisons to observed datasets for future work.

\section{Conclusions}\label{conclusion}

We investigated how feedback affects the mass distribution of gas structures in dwarf galaxy simulations. Our simulation suite included simulations repeated from the same initial conditions ($M_{\rm g}=7\times10^{7}\,{\rm M}_{\odot}$ and $M_{\star}=10^{7}\,{\rm M}_{\odot}$ embedded in a $10^{10}\,{\rm M}_{\odot}$ \citet{navarro1997} dark matter halo), but with different sources of feedback included (SNe, SNe+wind, SNe+radiation, SNe+winds+radiation). Gas structures were selected by isodensity contours at densities $10$ (cloud complex), $10^{1.5}$ (big clouds), and $10^{2}\,{\rm cm}^{-3}$ (clouds). For clouds, the mass distribution shows no feedback dependence, with each simulation presenting a similar steeply decreasing mass distribution. In contrast, for cloud complexes, the presence of radiation leads to a more top-heavy mass distribution, with other forms of feedback showing no impact on the distribution. 

We compare this with the mass distribution of star clusters in the same simulations \citepalias{andersson24starclusters} where earlier feedback, in particular radiation, was shown to steepen the function (i.e., opposite of what we find for cloud complexes; see Fig.~\ref{fig:cluster}). Since feedback did not affect the velocity dispersion distribution of clouds, and only ionizing radiation was found to affect the thermodynamics of the gas, increasing the temperature of the cold, dense gas phase, we conclude that the shape of the star cluster mass function must be determined by a combination of intra-cloud feedback regulation of star formation and, in the case of radiation, effects on the temperature of the parent gas clouds. The impact of radiation on gas temperature may also explain the generation of a larger number of massive cloud complexes, since the Jeans mass depends on the 3/2 power of the temperature. Finally, since feedback impacts the thermodynamics of the gas, but not the cloud mass distribution, our results suggest that the distribution of cloud masses is predominantly shaped by gravity. 

\begin{acknowledgements}
We thank the anonymous referee for insightful comments that led to significant improvements in this paper. LVQ was supported by the Science Pathways Scholars Program of Barnard
College, generously funded by Laura and Lloyd Blankfein. EPA and M-MML acknowledge support from
US National Science Foundation grant 
AST23-07950 and NASA Astrophysical Theory Program grant 80NSSC24K0935. EPA acknowledges resources from SNIC 2022/6-76 (storage) and LU 2022/2-15 (computing) for executing the simulations analyzed here.
\end{acknowledgements}
\bibliographystyle{aa}
\bibliography{ref}

\begin{figure}[tbh]
    \centering
    \includegraphics[width=0.95\linewidth]{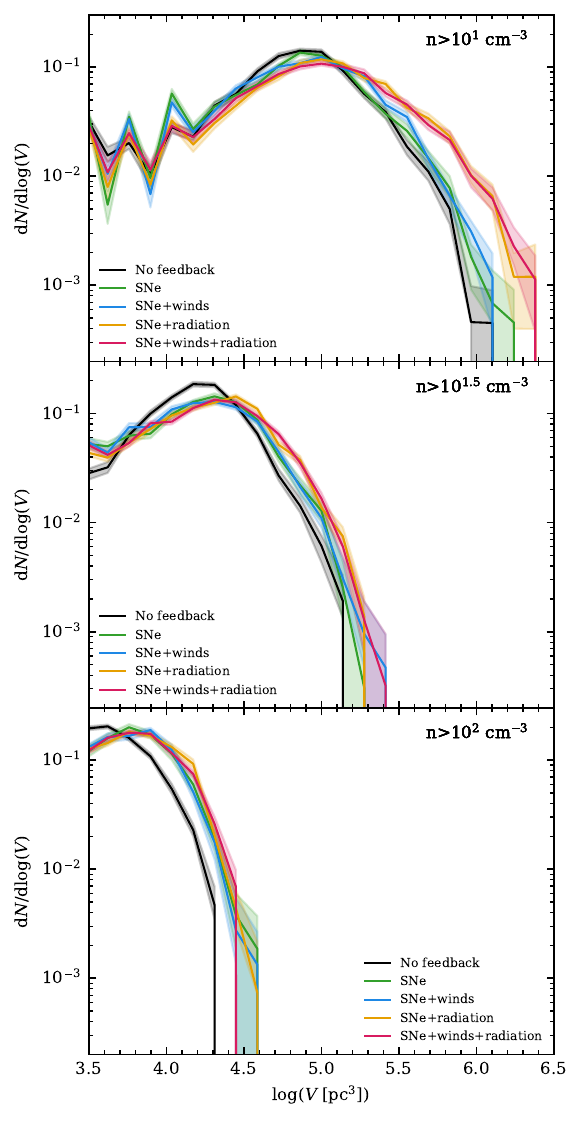}
    \caption{ 
       Volume distributions for cloud complexes with $n>10$~cm$^{-3}$ (top), big clouds with $n>10^{1.5}$~cm$^{-3}$ (middle), and clouds with $n>10^2$~cm$^{-3}$ (bottom). The lines show the median values, while the shaded regions show the 16th--84th percentile range.} 
    \label{fig:volfunction}
\end{figure}

\begin{figure}[tbh]
    \centering
    \includegraphics[width=0.95\linewidth]{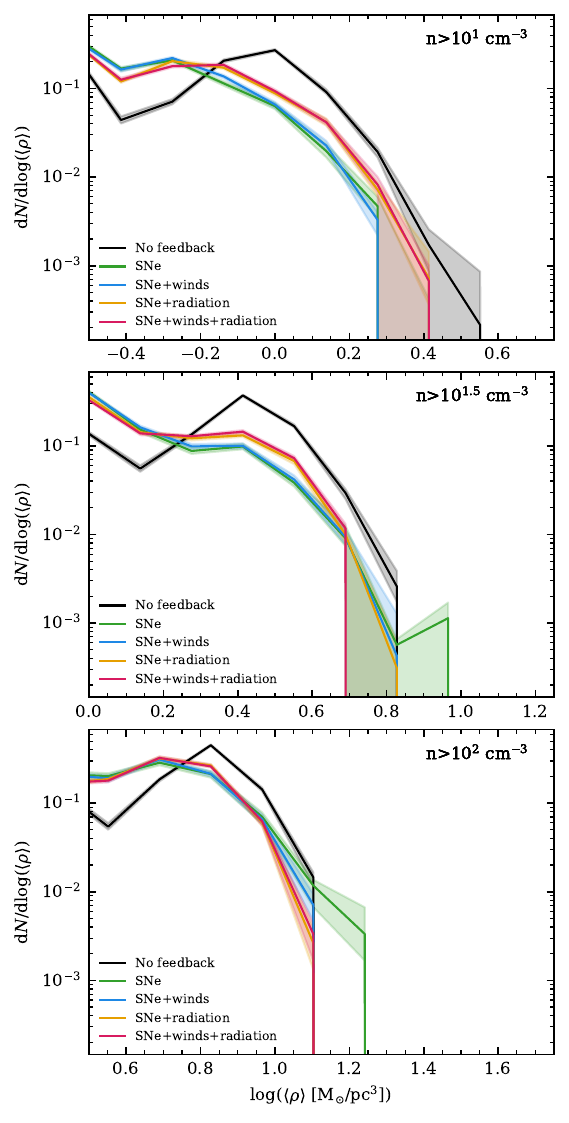}
    \caption{ 
       Distributions of mean density for cloud complexes with $n>10$~cm$^{-3}$ (top), big clouds with $n>10^{1.5}$~cm$^{-3}$ (middle), and clouds with $n>10^2$~cm$^{-3}$ (bottom). The lines show the median values, while the shaded regions show the 16th--84th percentile range. Note that the limits on the horizontal axis are different in the panels.} 
    \label{fig:rhofunction}
\end{figure}

\begin{appendix}
\twocolumn[{
\section{Volume distributions}\label{appendix}

\begin{multicols}{2}
We here demonstrate that the distributions of cloud volumes (Figure~\ref{fig:volfunction}) behave qualitatively similarly to the distribution of cloud masses (Figure~\ref{fig:massfunction}) discussed in Section~\ref{results}.  In both cases, prompt (winds and radiation) feedback leads to larger cloud complexes, but similar distributions of big clouds and clouds as delayed feedback (SN only) models. No feedback results in lower mass or volume clouds.  

However, the distributions of cloud average densities (Figure~\ref{fig:rhofunction}) behave somewhat differently than the masses or volumes, in that the model with no feedback produces objects with the highest average densities at all scales, while prompt feedback only makes a small difference for cloud complexes.
\end{multicols}
}]

\end{appendix}

\end{document}